\documentclass[aps,nofootinbib,twocolumn]{revtex4}
\usepackage{amsfonts,amssymb,amsmath}            
\usepackage[]{graphics,graphicx,epsfig}            
\usepackage{amsthm,bbold}

\newcommand{\ket}[1]{\left | #1 \right\rangle}
\newcommand{\bra}[1]{\left \langle #1 \right |}
\newcommand{\half}{\mbox{$\textstyle \frac{1}{2}$}}

\newcommand{\braket}[2]{\left\langle #1|#2\right\rangle}
\newcommand{\proj}[1]{\ket{#1}\bra{#1}}
\newcommand{\identity}{\mathbb{1}}
\renewcommand{\epsilon}{\varepsilon}
\newcounter{pp}

\begin{document}

\title{Comment on Partial Adiabatic Quantum Search}
\date{\today}
\author{Alastair \surname{Kay}}
\affiliation{Department of Mathematics, Royal Holloway University of London, Egham, Surrey, TW20 0EX, UK}
\begin{abstract}
The partial adiabatic search algorithm was introduced in \cite{1} as a modification of the usual adiabatic algorithm for quantum search with the idea that most of the interesting computation only happens over a very short range of the adiabatic path. By focussing on that restricted range, one can potentially gain advantage by reducing the control requirements on the system, enabling a uniform rate of evolution. In this comment, we point out an oversight in the original work \cite{1} that invalidates its proof. However, the argument can be corrected, and the calculations in \cite{1} are then sufficient to show that the scheme still works. Nevertheless, subsequent works \cite{2,3,4,5,6} cannot all be recovered in the same way.
\end{abstract}
\maketitle

In a series of recent papers \cite{1,2,3,4,5,6}, a number of authors have studied the algorithm of partial adiabatic quantum search. The idea of this algorithm is to only perform an adiabatic algorithm for quantum search in the neighbourhood of the critical point (vanishing gap). In this region, the Hamiltonian need only be varied at a constant rate, and, up to a constant factor, this is the optimal path through the region. While the basic premise is valid, existing analyses are insufficient, failing to take into account the potentially destructive interference effects of excited states. Here we point out where the failure in the reasoning occurs, and show how it can be strengthened so that, at least in special cases, the original claims of \cite{1} still hold. However, not all of the subsequent results in \cite{2,3,4,5,6} can be fixed.

In the quantum search algorithm \cite{search}, we consider a set of $N$ states, $\ket{x}$, of which some unknown subset, $S$, are marked ($|S|=M$). It is our task to find an instance of $S$ given that we can recognise it as such when we have it, which we do by slowly interpolating a system between two Hamiltonians
$$
H(\mu)=(1-\mu)H_s+\mu H_t,
$$
having prepared the system initially in the ground state of $H_s$ such that, when $\mu=1$, we produce the ground state of $H_t$. In \cite{1}, an arbitrary $H_s$ was allowed, while $H_t$ was restricted to being of projector form $H_t=-\proj{t}$. In order to show how and why the previous analyses fail, it is sufficient to restrict the form of $H_s$ to also being a projector, $H_s=\identity-\proj{\xi}$. In trade, we are more easily able to consider the case of searching for multiple items rather than just 1, which is pertinent to the later works, by replacing
$$
H_t=-\sum_{x\in S}\proj{x}.
$$
Having done so, we define the states
\begin{eqnarray*}
\ket{\xi}&=&\frac{1}{\sqrt{N}}\sum_x\ket{x}=\sqrt{\frac{N-M}{N}}\ket{\alpha}+\sqrt{\frac{M}{N}}\ket{\beta}	\\
\ket{\alpha}&=&\frac{1}{\sqrt{N-M}}\sum_{x\notin S}\ket{x}	\\
\ket{\beta}&=&\frac{1}{\sqrt{M}}\sum_{x\in S}\ket{x}.
\end{eqnarray*}
In order to ensure that the correct output is produced with high fidelity, the evolution must be sufficiently slow, which is based on the criterion
$$
\left|\bra{\lambda_{GS}(t)}\frac{dH}{dt}\ket{\lambda_{GS}(t)}\right|\leq \epsilon\Delta(t)^2
$$
where $\Delta(t)$ is the instantaneous energy gap between the ground state and the first excited state, $\ket{\lambda_{GS}(t)}$ is the ground state, and $0<\epsilon\ll 1$ is an error parameter.

The energy gap and ground state can both be calculated analytically in this case because the subspace $\{\ket{\alpha},\ket{\beta}\}$ is preserved, meaning we just have to analyse a $2\times 2$ matrix \cite{adiabatic, localsearch}. This allows one to calculate the function $\mu(t)$ and prove its optimality \cite{localsearch}. We will not repeat the full details of these calculations here. However, we will point out an important symmetry property. Define a unitary $U$ such that
\begin{eqnarray*}
U\ket{\beta}&=&\ket{\xi}	\\
U\ket{\alpha}&=&-\sqrt{\frac{M}{N}}\ket{\alpha}+\sqrt{\frac{N-M}{N}}\ket{\beta}.
\end{eqnarray*}
This means that $U\ket{\xi}=\ket{\beta}$ (and $U=U^\dagger$), allowing one to show that
$$
UH(\mu)U^\dagger=H(1-\mu),
$$
i.e.\ there is a symmetry about the point $\mu=\half$.

The partial adiabatic quantum search was proposed in order to avoid the requirement of varying $\mu(t)$ in time, while still gaining the square-root speed-up of the quantum search (if one uses a constant $\frac{d\mu}{dt}$, slow enough to fulfil the adiabatic criterion everywhere, then that square-root speed-up is lost). The protocol involves:
\begin{enumerate}
\item Prepare the system in the state $\ket{\xi}$.
\item Evolve the Hamiltonian between two points $\mu_-=\half-\delta$ and $\mu_+=\half+\delta$ for some $\delta$ (this has typically been specified as $1/(2\sqrt{N})$, although we will leave it unspecified for now).
\item Measure the system in the computational basis and decide if the search was successful. If not, run again.
\end{enumerate}
For some probability of success $p(\delta)$ for a single run, and assuming constant $\frac{d\mu}{dt}$, the expected running time of the algorithm would be
$$
\frac{2\delta}{p\frac{d\mu}{dt}},
$$
just leaving one to calculate the probability $p$. This is where the flaw arises in previous analyses (for example, just before Eq.\ (29) in \cite{1}, where the combined success probability is given). It has been claimed that the success probability is
$$
p=|\braket{\xi}{\lambda_{GS}(\mu_-)}|^2 |\braket{\beta}{\lambda_{GS}(\mu_+)}|^2.
$$
At first glance, this looks vaguely plausible -- if the initial ground state at $\mu_-$ is mapped to the final ground state at $\mu_+$ by the adiabatic evolution, then we're interested in the probability that our initial state was in the ground state ($|\braket{\xi}{\lambda_{GS}(\mu_-)}|^2$) and that the final state is in the target state ($|\braket{\beta}{\lambda_{GS}(\mu_+)}|^2$). However, this analysis is only valid if one were to project onto the ground state space, which is not what happens.

To emphasise this, consider the limit as $\delta\rightarrow 0$. We {\em know} that all that happens in the described protocol is that the ground state $\ket{\xi}$ is prepared, and it is measured in the computational basis, and then verified to be in the subspace $S$ (or not). This happens with exactly the probability
$$
\sum_{x\in S}|\braket{x}{\xi}|^2=\frac{M}{N},
$$
which requires $O(N/M)$ trials to successfully find a good $x$. In comparison, the previous argument claims the success probability must be
$$
p=|\braket{\xi}{\lambda_{GS}(\half)}|^2 |\braket{\beta}{\lambda_{GS}(\half)}|^2.
$$
where $|\braket{\xi}{\lambda_{GS}(\half)}|^2=|\braket{\beta}{\lambda_{GS}(\half)}|^2=\half+\half\sqrt{\frac{M}{N}}$. Hence, it suggests $p>\frac{1}{4}$ and the algorithm only has to be run a constant number of times!

The problem is that by not projecting onto the ground state space, you can get interference from the excited states. Here's the most extreme example: Prepare an initial state $\ket{0}$, where the target state is $\ket{1}$ using the intermediate state $(\ket{0}+\ket{1})/\sqrt{2}$. By the previous arguments, $p=\frac{1}{4}$, while the true success probability is 0.

How should the calculation be corrected? If one does not want to explicitly consider the interference term (after all, we do not want to describe how the excited states behave in an adiabatic algorithm), one must assume that the interference term is maximally destructive. Starting with the state $\ket{\xi}$, we write it as $a\ket{\lambda_{GS}(\mu_-)}+\sqrt{1-|a|^2}\ket{\lambda_-^\perp}$ where $a=\braket{\lambda_{GS}(\mu_-)}{\xi}$ and $\ket{\lambda_-^\perp}$ is orthogonal to the ground state at $\mu_-$. Under the adiabatic evolution (which we assume to be perfect here), this becomes
$a\ket{\lambda_{GS}(\mu_+)}+\sqrt{1-|a|^2}\ket{\lambda_+^\perp}$, and the probability of finding this in the state $\ket{\beta}$ is
$$
p=\left||a|^2+\sqrt{1-|a|^2}\braket{\beta}{\lambda_+^\perp}\right|^2.
$$
Here we have used the symmetry property that shows
$$
\braket{\lambda_{GS}(\mu_+)}{\beta}=\bra{\lambda_{GS}(\mu_-)}U^\dagger\ket{\beta}=a.
$$
Given that we don't want to assume anything about the evolution of the excited states, we parametrise $\braket{\beta}{\lambda_+^\perp}=\chi e^{i\phi}$ with $0\leq \chi\leq\sqrt{1-|a|^2}$. Hence
\begin{eqnarray*}
p&=&|a|^4+(1-|a|^2)\chi^2+2|a|^2\sqrt{1-|a|^2}\chi\cos\phi	\\
&\geq &(|a|^2-\sqrt{1-|a|^2}\chi)^2.
\end{eqnarray*}
Overall, this leaves us with the bound
$$
p\geq\left(\max(0,2|a|^2-1)\right)^2
$$
We now see that it is insufficient to bound the overlap of the initial state with the ground state by a constant, as was claimed in \cite{1,2,3,4,5,6}, but it must be bounded over $1/\sqrt{2}$ by a constant factor.

For the special case of projector Hamiltonians, we can see that the values of $\mu_{\pm}$ suggested in \cite{1} are, in fact, acceptable. Consider
$$
\delta=\gamma\sqrt{\frac{M}{N-M}}.
$$
This gives
$$
|a|^2=\half+\frac{\sqrt{\frac{M}{N}}+2\sqrt{1-\frac{M}{N}}\gamma}{2\sqrt{1+4\gamma^2}},
$$
which, to leading order in $M/N$, gives a success probability of
$$
p=\frac{4\gamma^2}{1+4\gamma^2}.
$$

In order to achieve adiabaticity, we require
$$
\frac{d\mu}{dt}\leq\frac{(1-4\mu(1-\mu)(1-\frac{M}{N}))^{3/2}}{(1-\frac{M}{N})(1-2\mu)}\epsilon.
$$
There are now two options. For the optimal path, we would integrate this relation between $\mu_-$ and $\mu_+$, finding a total evolution time for one shot of
\begin{equation}
T=\frac{1}{\epsilon}\sqrt{\frac{N}{M}}\left(1-\frac{1}{\sqrt{4\gamma^2+1}}\right). \label{eqn:optimal}
\end{equation}
Alternatively, we can find the time for which $\frac{d\mu}{dt}$ is maximum, and use that as the fixed rate of evolution (as is the idea behind the partial adiabatic search) over the period $\delta \mu=\mu_+-\mu_-=2\delta$. If $\gamma\leq\frac{1}{2\sqrt{2}}$, then this maximum rate occurs at the time $\mu_+$. Otherwise, it occurs at $\mu=\half+\frac{1}{2\sqrt{2}}$, yielding a running time for one shot of
\begin{equation}
T=\left\{\begin{array}{cc}
\frac{1}{\epsilon}\sqrt{\frac{N}{M}}\frac{4\gamma^2}{(1+4\gamma^2)^{3/2}}	& \gamma\leq\frac{1}{2\sqrt{2}}	\\
\frac{1}{\epsilon}\frac{4\gamma}{3\sqrt{3}}\sqrt{\frac{N}{M}}	& \gamma>\frac{1}{2\sqrt{2}}
\end{array}\right.	\label{eqn:constant}
\end{equation}
For large $\gamma$ (such that, with high probability, only one run is required), there is a constant overhead of approximately $4\gamma/3\sqrt{3}$ in running time compared to the optimal path. On the other hand, the minimum expected running time of $T=\frac{4}{3\sqrt{3}\epsilon}\sqrt{\frac{N}{M}}$ occurs for $\gamma=\half$. Nevertheless, in all cases the running time is $O(\sqrt{N/M})$, matching the circuit-based complexity for the problem.

In fact, the original calculations in \cite{1} are also strong enough to argue that for the case of search with one marked item, but for a general $H_s$, the success probability is finite. However, subsequent papers have not been so fortunate, and this has caused a certain degree of confusion. The presentation here is sufficient to indicate how, if possible, these results may be remedied, by proving stronger bounds on the overlaps, $a$.

\section{Optimality}

There are several parameters that one can try to optimise over with this algorithm in order to minimise the expected running time -- the starting point $\mu_-$, the finish point $\mu_+$ and the rate $\frac{d\mu}{dt}$ at which one moves along the path. By symmetry, if we find an optimal $\mu_-$ then the optimal $\mu_+$ is given by $\mu_+=1-\mu_-$. This was already incorporated into our analysis, and we then showed that, for the variant where $\frac{d\mu}{dt}$ is held constant, the optimal choice is $\mu_-=\half-\half\sqrt{\frac{M}{N-M}}$.

As for the optimal path $\mu(t)$ between any two points $(\mu_-,\mu_+)$, we already know the optimal path for $(0,1)$ \cite{localsearch}. For any two points $0\leq \mu_-<\mu_+\leq 1$ we could think about using the relevant section of that path. This is exactly what we calculated for Eq.\ (\ref{eqn:optimal}). Moreover, this must be optimal -- if there were a better path between $(\mu_-,\mu_+)$, meaning it can be executed in less time, that shorter path could be substituted into the $(0,1)$ path, making it shorter. However, since it's optimal, it can't be shortened. We therefore conclude that the path for Eq.\ (\ref{eqn:optimal}) is optimal for any $(\mu_-,\mu_+)$ and, since the path for Eq.\ (\ref{eqn:constant}) yields an evolution time that is only a constant factor different, it also possesses the optimal scaling.

\section{Conclusion}

In this comment, we have argued that the criterion used in \cite{1,2,3,4,5,6} for assessing the success probability of the partial adiabatic quantum search algorithm was incorrect, and had the potential to seriously overestimate the probability of success. We have described how to correct for this and shown that the partial adiabatic quantum search algorithm is still feasible.

We would like to end with a word of caution, however. We have worked in the regime where the success probability is high, so the expected number of runs of the algorithm is small. In principle the same analysis can also be applied in the limit where the success probability is very small. However, a subtle modification is required because, to date, we have assumed that the cost of recognising a valid solution is negligible. This would no longer be true, and one must assume that each such test takes a constant time.

\end{document}